%Paper: astro-ph/9305008
%From: DCH researcher <mirek@mathematics.edinburgh.ac.uk>
%Date: Fri, 7 May 93 14:29:37 BST

\def\gtorder{\mathrel{\raise.3ex\hbox{$>$}\mkern-14mu
             \lower0.6ex\hbox{$\sim$}}}
\def\ltorder{\mathrel{\raise.3ex\hbox{$<$}\mkern-14mu
             \lower0.6ex\hbox{$\sim$}}}
\def\etal{{\sl et al.}~}
\font\BF=cmbx10 scaled \magstep1
\magnification=\magstep1
\null\bigskip
\centerline{{\BF STATISTICS OF N-BODY SIMULATIONS}}
\medskip
\centerline{{\BF I. EQUAL MASSES BEFORE CORE COLLAPSE}}
\bigskip
\centerline{Mirek Giersz\footnote{$\!{}^1$}
{On leave from the N.  Copernicus
Astronomical Centre, Bartycka 12, 00-716 Warsaw, Poland}
and Douglas C.  Heggie}
\medskip
\centerline{Department of Mathematics and Statistics,}
\centerline{University of Edinburgh,}
\centerline{King's Buildings,}
\centerline{Edinburgh EH9 3JZ,}
\centerline{U.K.}
\null\bigskip
{\bf ABSTRACT}

{\parindent=0pt
We study the dynamical evolution of idealised stellar systems by
averaging results from many $N$-body simulations, each having modest
numbers of stars. For isolated systems with stars of uniform mass, we
discuss aspects of evolution up to the point of core collapse:
relaxation and its $N$-dependence, the evolution of the density profile,
the development of the velocity dispersion and anisotropy, and the rate
of stellar escape.}

\bigskip


\vfill\eject

{\bf 1 \quad INTRODUCTION: THE STATISTICAL STUDY OF $N$-BODY SYSTEMS}
\medskip

There have been dramatic improvements in the performance of
astrophysical $N$-body simulations since the pioneering work of von
Hoerner (1960). Several, e.g. regularisation of close encounters, have
taken place in the domain of software, i.e.  developments of the
programs used. Many of these are documented in Aarseth (1985), and
indeed due to him, but others are beginning to have an impact on the
field of collisional stellar dynamics, including tree codes (Barnes \&
Hut 1986, McMillan \& Aarseth 1993)

Of comparable importance are improvements in hardware, many of which
have been exploited quickly into collisional stellar dynamics. Among
these are special-purpose devices such as Grape-2 and its descendants
(Ito et al. 1991), and general-purpose machines such as the Cray Series
supercomputers.  The latter have, for example, been used effectively by
McMillan, Hut \& Makino (1990) in their studies of systems with primordial
binaries. Even so, dedicated workstations are still competitive (Aarseth
\& Heggie 1992).

A further group of hardware devices suitable for stellar dynamics are
parallel processors, such as the Connection Machine (Hut \& Makino 1989).
Other machines
which have been used for problems in stellar dynamics include the
Distributed Array Processor, which was used in a study of gravothermal
behaviour (Heggie 1989), and transputer arrays (Sweatman 1993a). Similar
machines with faster (vector) processors are now available, and have
been used in the project described later in this paper.

The improvement in computational power over the last three decades can
be exploited in several different ways. One aim is to study larger
systems.  This is driven by the fact that real globular clusters have
many more stars than the largest useful computations, by at least an
order of magnitude.  It is not at all clear that this is necessarily the
best way of using increased computing power, as the discussion of the
following section shows.

Another aim is simply to carry out calculations more quickly.  For
example, workstation calculations on systems with $2500$ stars and a
substantial proportion of primordial binaries may still require as much
as $2000$ cpu hours (Aarseth \& Heggie 1992).

A third aim is to exploit the power to produce better results, which
means better statistics. It is well known that little can be done to
improve the accuracy of the positions and velocities of the stars:
errors in these quantities grow exponentially on a time scale which is
about $t_{cr}/8$ (Goodman, Heggie \& Hut  1993), where $t_{cr}$ is the crossing
time.  Therefore, on the timescales for interesting dynamical evolution
(which are of order $10^2 t_{cr}$ for $N = 1000$) the positions
and velocities of the stars are quite wrong.  Nevertheless it is an
article of faith among practitioners of $N$-body modelling that the {\sl
statistical} results are meaningful (e.g. Aarseth \& Lecar 1975).  Therefore,
when we speak of improving the results, we mean improving the quality of
statistical results.  Furthermore, since statistical results are the
only useful results from these simulations, it is surprising that in
many cases we are content with a single measurement of the statistical
results of importance.  Of course, it may be that measurements taken at
different times on the same system will allow a certain amount of
statistical sampling, but it is by no means clear how independent these
are. It is much safer, as stressed, for example, by McMillan, Hut \& Makino
(1991), to
compute several simulations with identically distributed initial
parameters, differing only in the random numbers used to compute each
realisation.

This approach has been exploited previously by Casertano (1985), who
studied aspects such as the distribution of escape energies, though for
systems with 8 stars, and by McMillan, Caseranto \& Hut (1988), who
concentrated on
an intensive study of the relaxation timescale in $N$-body systems with
$N$ in the range $16<N<1024$.  Our emphasis is on the wider range of
dynamical processes, not excluding relaxation, which are of importance
in astrophysical applications, such as the evolution of the spatial
distribution of stars, anisotropy, escape, binary activity, and so on.

This paper continues with a summary of the software and hardware issues
which are relevant to surveys of the kind that we have attempted.  We
then discuss a number of results for isolated systems in which all stars
have the same mass: the evolution of the space distribution of the stars
and the distribution of velocities (including anisotropy), and the rate
of escape.  In this paper we present and discuss data up
to the end of core collapse.  The next paper in this series continues
the discussion well into the post-collapse regime, and adds some
detailed results on the evolution of binaries, and their effects on the
system.  Subsequent papers will deal with systems of stars in which the
dynamics is more complicated in various ways: systems with a spectrum of
masses, tidally truncated systems, and systems affected by mass-loss
through stellar evolution.
Further papers
will also consider the effects of
rotation, which are less easily modeled by Fokker-Planck techniques,
and mass segregation.

One purpose of these studies is to test the results of the valuable
paper by  Chernoff \& Weinberg (1990), who studied a similar range of
problems (up to the
point of core collapse) with the Fokker-Planck method.  More generally
we aim to compare our $N$-body results with those of a variety of simple
models and scaling arguments. It might be argued {\sl ab initio} that
such an attempt is either bound to fail or to lack meaning, since these
simple models usually apply in the limit of large $N$, and it is not
clear whether the small models we study lie sufficiently far into the
asymptotic regime.  But in cases where the $N$-body results do agree
with the simple models, we can have some confidence that the simple
models are an adequate description of the phenomenon in question and its
underlying mechanisms.  It might also be said that we take the approach
of a theorist attempting to devise a model for a set of observations:
starting from familiar physical principles or models (in this case,
those of stellar dynamics), our aim is to construct the simplest
consistent theory which accounts for the observations. We allow
ourselves to adjust ill-determined parameters to optimise the fit, and
include no detailed features which are not required by the data.

\bigskip
{\bf 2 \quad A PARALLEL STUDY OF $N$-BODY STATISTICS}
\medskip
{\bf 2.1 \quad Hardware}
\medskip

Much of the computational work reported below has been carried out on two
different computers, both installed at the Edinburgh Parallel Computing
Centre.  One is the Meiko transputer array, the other the ``Grand
Challenge Machine'', also called maxwell, which is an array of $64$ i860
processors, installed in the autumn of 1990.  Though parallel computers
are often thought of as being very hard to program, it is quite easy to
mount a large number of independent $N$-body problems, as the only
messages passed between processors involve simple tasks of assigning
input parameters and gathering output.  The difficulties are not much
greater than those involved in mounting an $N$-body program on any new
single-processor hardware.

The transputer array (ECS, for Edinburgh Computing Surface) contains
$400$ processors, each of
which has communications and at least $4$Mbytes of local storage.
As the machine
is configured for multi-user operation, generally only smaller ``domains'',
of up
to $131$ transputers, are available to each user.  Programs are developed
and run under a software environment known as ``CS-Tools''.  Individual
processors are rated at about $1$Mflop.  In practice, the code NBODY1
(Aarseth 1985)
runs at about one fifth times the speed obtained on a Sun ELC.

The Grand Challenge Machine has a similar structure and environment,
which made the task of porting code to this machine very quick and
painless.  It differs mainly in the speed of the processors, which are
roughly 100 times faster than those of the ECS.  Each i860 is a vector
processor, with communications and local storage, which is capable in
principle of a performance of $10^2$ Mflops, and the entire machine has
a peak performance of about $5$Gflops.  It is also very cost-effective,
at around {\sl\$}1M.  The main purpose of the machine is computations in
high-energy physics, and performance approaching the peak has been
obtained in this application.  But such performance is only available
with suitable hand-crafted code, and much more modest performance is
obtained in routine FORTRAN programs; NBODY1 runs at a speed about 5 times
faster than on the ECS. On the other hand, while it was
being installed and tested (in the winter of 1990/91), much computing time
was available, and was
used for some of the calculations described below (see \S2.4).

Later in the course of this project a DEC Alpha superscalar machine
(called ``fringe'') was
installed on field test at the University of Edinburgh. This is a
single-processor machine, but very fast, and data from large numbers of
$N$-body simulations were built up sequentially.
\bigskip

{\bf 2.2 \quad Statistical considerations}
\medskip
As already mentioned in \S1, there are at least two ways in which the
power of parallel computers can be exploited, either in larger
simulations or in more numerous simulations.  In order to assess the
value of these two approaches let us consider the i860 machine, with 64
processors, and let us  suppose that it is feasible to study a 1000-body
problem on a single processor.

First we consider what can be achieved by using the entire machine for a
single large calculation.  With $64$ processors it would be feasible to
study a problem in which the computational effort was $64$ times larger
(provided that the system could be used at $100\%$ efficiency).  With a
simple direct summation algorithm, however, the computational effort
grows roughly as $N^3$ (Aarseth 1985, with an extra power of $N$ because
we are concerned with processes occurring on the relaxation time scale),
and so it is feasible actually to compute a 4000-body system.  Assuming
that statistical results yield estimates with standard deviation
proportional to $N^{1/2}$, this means that the signal-to-noise ratio for
these results is improved by a factor of $2$, compared with what can be
achieved in a comparable time with a single processor.

Now suppose the same machine is used to run $64$ simultaneous (but
independent) $1000$-body simulations (and it is very easy to do so at
virtually $100\%$ efficiency).  Then the signal-to-noise ratio is improved
by a factor of $8$ compared with results for a single simulation.
Therefore it is clear that the statistical quality of the results is
optimised by adopting the second strategy.  The exact figures, and
especially the assumed number of bodies for a feasible single-processor
simulation, are irrelevant.  It is easy to see, assuming $100\%$
efficiency, that the second strategy is better provided that the
computational effort varies as $N^\alpha$ for any $\alpha>1$.  For
Ahmad-Cohen schemes the empirical value of $\alpha$ is of order $2.6$
(Aarseth 1985, modified as before),
while a tree-based code yields $\alpha\geq 2$.

Before we conclude that it is always better to run large numbers of
modest simulations than one large one, we should consider other issues
than simply statistical quality.  There are some phenomena which may
well behave very differently in small and large systems - gravothermal
effects, which manifest themselves only for systems containing many
thousands of particle (Goodman 1987) are one.  Another phenomenon which
would be rather difficult to study by the use of large numbers of small
simulations is the phenomenon of core motions (Makino \& Sugimoto 1987,
Heggie 1989, Sweatman 1993b),
because such motions could not be expected to occur coherently in
different systems.  The statistical analysis required would be rather
different from the kind that is considered in this paper.  Possibly the
estimation of autocorrelation functions from a large number of
simulations would be a fruitful approach, but it is not pursued here.
Nevertheless, for many phenomena of interest, such as those studied in
this paper, there is no known qualitative difference between large and
small systems, and the most scientifically productive way of exploiting
parallelism is to run large numbers of simulations in parallel.

\medskip
{\bf 2.3 \quad Software}
\medskip

In this project some pilot sets of calculations were computed using a
version of Aarseth's code NBODY1 (Aarseth 1985), which uses individual
time steps.  This code is quite effective up to the point where binary
activity becomes a problem, and was used in a preliminary study of core
collapse.  Subsequently, a code including regularisation of binary
encounters was adopted, along the lines described briefly in Heggie
(1973), but extended to include ``freezing'' of hard isolated binaries
(Aarseth 1985).  This allowed the computation of new series of isolated
models, with $250$, $500$ and $1000$ equal-mass stars in each, which
included the end
of core collapse and subsequent reexpansion.  This code was still too
inefficient to deal with very hard interactions between stars and binaries
and binaries themselves during the advanced collapse and post-collapse phases.
Some cases were stopped because of the occurrence of excessive energy errors,
and this resulted in substantial
deterioration of overall statistics. Therefore our definitive sets of
calculations
of isolated systems with $250$, $500$, $1000$ and $2000$ equal-mass stars
were carried
out with Aarseth's code NBODY5.  It was also adopted, with some modifications,
for calculations including tidal effects, mass loss through stellar
evolution, a mass spectrum, etc., which will be described in later
papers in this series.

One general point to made about software in a parallel environment is
reliability. On a scalar machine it is possible, if irksome, to give
individual attention to a model when some unforeseen combination of
circumstances, or a programming error, leads to a serious degradation in
accuracy. This is not practical if many models are computed
simultaneously. Then it is necessary to write into the program
instructions for automatically restarting a computation from a
previously stored configuration, but with new parameters which will
allow a more accurate integration (McMillan et al. 1991).

For the organisation of parallel computations some additional
programming considerations are relevant.  The domain of processors is
conceptually divided into a single `master' and a number of `slaves', one
for each
simulation in the set.  The computation is initialised when the master
issues initial data (e.g. $N$, and a seed for a random number generator)
to each slave.  Each slave runs a version of an $N$-body program.  When
the slave is ready to output analyzed data (which in our calculations was done
every time unit) it does so to the master, which adds each such line of
output to a single results file.  Each line is identified with the time
of the output (in $N$-body time units) and the identity of the slave.
The calculations running on different slaves proceed at different rates,
and so the lines are not in any definite time-order. Nevertheless, after
all slaves have finished their simulations, the results file can be
analyzed, all lines of output which refer to the same time (but from
different slaves and simulations) being combined statistically.

Each slave also occasionally produces a file of data which is sufficient
to restart the run, if this is required because of a degradation in
accuracy, or if the run has been interrupted for any reason, or has to
be scheduled over several production periods (in a shared environment).
On the other hand this was not regarded as part of the ``scientific''
output of the project, as these data sets were regularly overwritten.

\medskip
{\bf 2.4 \quad Initial conditions and data}
\medskip

In this section we discuss the initial parameters of the runs, summarise
the data which was output, and present some statistics on the sets of
runs which have been completed and are relevant to the results of this
paper.

The initial conditions for these models were drawn from a Plummer model,
with equal masses, and no primordial binaries (except for any which
occur by accident).  The systems are isolated, with no external tidal
effects.  Units, which may be referred to below as ``$N$-body units'',
are standard (Heggie \& Mathieu 1986), and lead to the
following initial values: the crossing time  is $2\surd2$, the virial
radius is 1, and the initial rms 3-dimensional speed is $1/\surd2$.
These units are achieved by first shifting to the barycentric frame and
then rescaling the positions and velocities so that the initial kinetic
and potential energies take the required values.

At each time unit, each model outputted a list of data.  For these
systems with equal masses it was not felt necessary to store data on
individual particles, and so most of the items are statistical.  To some
extent the list evolved throughout the course of our project, as new
aspects of interest emerged.  The following list is comprehensive, but
for the reason just stated, not all of this information is available for
all sets of calculations.

\noindent(1) the number of the run in the current series;

\noindent(2) the time in $N$-body units;

\noindent(3) the total mass of non-escapers (defined in \S3.3);

\noindent(4) the virial ratio (i.e. the ratio of kinetic to binding
energy of all bound stars, excluding ``internal'' kinetic and potential
energies of recognised binaries);

\noindent(5) the central potential (i.e. the potential at the density centre);

\noindent(6) the core radius $r_c$ (defined as in Casertano \& Hut
(1985)).  In addition, the core radius was also computed from the
definition $r_c^2=3v_c^2/4\pi G\rho_c$, where $v_c$ and $\rho_c$ are the
central three dimensional velocity dispersion and central density,
which in turn were computed from information on stars within the first
Lagrangian radius (cf.(13) and (31) below);

\noindent(7) the mass of the core, computed in two ways: (i) as in
Casertano \& Hut (1985), and (ii) by summing the masses of stars
within a sphere of radius $r_c$ (calculated by the second method
outlined in (6) above) centered on the density centre, which in turn
is defined below in
\S3.1.1);

\noindent(8) the ratio of the central density (defined as in Casertano \&
Hut (1985)) to the mean density within the half-mass radius;

\noindent(9) the number of stars in the core (computed in two ways, as in (7)
above);

\noindent(10) the three-dimensional root mean square speed of stars
in the core (except that a binary, defined here as a regularised bound
pair, contributes only the speed of motion of its barycentre);

\noindent(11) the coordinates and velocity components of the density
centre, the latter being defined by the change in coordinates over one
time unit;

\noindent(12) the coordinates and velocities of the centre of mass;

\noindent(13) Lagrangian radii (radii of spheres containing fixed
fractions of the mass of bound stars, relative to the density centre)
for fractions $1$, $2$, $5$, $10$, $20$, $30$, $40$, $50$, $75$ and $90\%$;

\noindent(14) the number of bound stars;

\noindent(15) the number of escaping single stars;

\noindent(16) the number of regularised binaries (the parameters which
determine whether a pair is regularised taking standard values
(Aarseth 1985); sometimes we loosely refer to such binaries as ``hard'');

\noindent(17) the number of unregularised binaries with energy greater roughly
than $0.1 kT$, where $kT$ is defined as $2/3$ of the current mean kinetic
energy of
the single stars and of the barycentres of binary stars);

\noindent(18) the number of {\sl merged} pairs (defined as in Aarseth
1985 with standard parameters);

\noindent(19) the number of {\sl triple} or {\sl quadruple} configurations
(defined as in (18) above);

\noindent(20) the total energy of escaping single stars;

\noindent(21) the total and maximum internal binding energy of all
unregularised binaries with energies above $0.1 kT$;

\noindent(22) the total and maximum internal binding energy of all
regularised binaries;

\noindent(23) the total energy of the centres of mass of escaping regularised
binaries;

\noindent(24) the total internal energy of escaping regularised binaries;

\noindent(25) the internal binding energy of {\sl triple} and {\sl quadruple}
configurations (defined as in (19) above);

\noindent(26) the internal binding energy of {\sl mergers} (defined as
in NBODY5);

\noindent(27) the total internal and external energy of the bound
members of the system, defined as in Aarseth \& Heggie (1992);

\noindent(28) the cumulative integration error in the total energy;

\noindent(29) the maximum and minimum internal energy and radius
(relative to the density centre) of all regularised binaries;

\noindent(30) the names of stars forming the hardest binary;

\noindent(31) mean square radial and tangential velocities of stars
between successive Lagrangian radii, and the same quantities corrected for
escaping stars which have not yet been removed from the $N$-body
calculation.

The output from the various models in each series were collected in a
single output file. For each series the typical size of an output file
is typically $35$Mbytes.  The file for each series was later processed so
as to produce statistics for all models in that series at each time.  As
each line is read, the time to which it refers is noted, and the data
{}from that line are added to the statistical data for that time.

Table 1 summarises both the extent of the computational effort in
these calculations and the numbers of particles and simulations.  Smaller
numbers of simulations with $N>1000$ have also been included for comparative
purposes.  Further discussion and results, including some data from a
$10000$-body simulation, are considered in a paper by Giersz \& Spurzem
(1993). It is worth pointing out that the statistical value of these
various sets of simulations are reasonably well balanced:
for example, one set consisting of $40$ simulations with $250$ bodies is
comparable with one simulation containing $10^4$ stars, as the total number of
particles is the same.

Before discussing the results of these simulations we mention here
some data about their accuracy, as measured by energy conservation.
Throughout the core collapse period the mean of the absolute value of
the total energy error is monotonically increasing and reaches at the
time of collapse values between $10^{-5}$ to $10^{-3.8}$ for different
$N$. As can be expected from simple theoretical arguments (e.g. the
larger time to core collapse for larger $N$) the smaller value is for
$N=250$ and the larger for $N=2000$. The energy error for individual
cases never exceeded $0.002$.

\bigskip
{\bf 3 \quad STATISTICAL RESULTS FROM $N$-BODY SIMULATIONS}

\medskip

{\bf 3.1 \quad Spatial evolution}
\medskip
{\bf 3.1.1 \quad Evolution of Lagrangian radii}
\medskip

We now discuss the results from these computations up to about the time
of core collapse, and begin here with the distribution of mass.  We
assume spherical symmetry relative to the ``density centre'' (computed
as in Aarseth \& Heggie 1992).  In what follows, ``Lagrangian radii" are
the radii of imaginary spheres containing a fixed fraction $f$ of the
$N$ bound stars in the system, i.e.  excluding escaping stars.  Where
this fraction corresponds to an integral number $fN$ of stars, the
radius is that of the $fN$th star in order of distance from the density
centre.  Where $fN$ is fractional, the value is interpolated.

Combining data from a large number of runs may be done in several ways,
and it is not clear at first how consistent the different results will
be, or what is the most effective method. One could, for example,
combine the data on individual stars from all cases at a given time, and
then compute the Lagrangian radii for this enlarged ensemble of stars.
Because of the limited data which we elected to output, however, this
method could not be adopted, and we chose instead a suitable statistical
estimator based on the set of Lagrangian radii obtained for each case
individually.  For example, Fig.1 shows data for the Lagrangian radius
defined by the innermost $5\%$ of the mass, for $N = 250$. These results
indicate that the mean and median (over all cases computed) behave very
similarly, and from now on we shall concentrate on the mean.

The improvement in the quality of the data obtained
by combining results from many cases in this way is illustrated in
Fig.2, which show the means for all Lagrangian radii in one set of
models, Fig.2a, and the results for a single, typical model, Fig.2b.
The improvement in
``signal-to-noise'' is comparable to what one would expect from simple
``root $n$'' considerations. It is this improvement which will allow us
to re-examine several quite old problems in collisional stellar dynamics
which have lain relatively fallow for want of results of sufficient
statistical quality.

\smallskip
{\bf 3.1.2 \quad $N$-dependence of dynamical evolution}
\medskip
The effect of the number of stars on the evolution of the Lagrangian
radii is illustrated in Fig.3.  The most obvious feature of this result
is the increase of the time scale for core evolution with $N$. This is
consistent with the assumption that the evolution is driven by two-body
relaxation, and that the relaxation time increases nearly linearly with
$N$ (cf. Spitzer 1987).  More precisely, the theoretical result is
that the relaxation time scale varies as $N/\ln(\gamma N)$, where
$\gamma$ is a constant.  In what follows we analyse this question
quantitatively in a number of ways, though the order in which these
are described differs from the order in which they were carried out.

One procedural point has to be mentioned first. Fig.4 shows, for $N =
250$, a comparison between the $N$-body results and two approximate
(continuum) models, which are based on the theory of relaxation and
described more precisely below. The point we wish to emphasise is the
fact that the $N$-body results lie {\sl initially} below those of the
continuum models, which correctly give the value for a Plummer model.
The reason for this discrepancy can be traced to several sources, some
of which arise naturally from the way in which the initial conditions
are scaled. For example, if $250$ point masses are spatially
distributed as in a Plummer model, the resulting potential will not be
exactly that of the underlying Plummer model, and it is probable that
the mean potential energy of a large number of simulations is biased
away from its true value.  Therefore, when the positions of the
particles are scaled to the desired value of the total potential,
there is a resulting bias in the spatial distribution. Another bias
arises from use of the potential centre, which tends to be a point of
overdensity in the system, and therefore Lagrangian radii tend to be
smaller than they should be. The different values of the initial $2\%$
Lagrangian radii for various $N$ are clearly visible on Fig.3.

Approximation correction for these biases has been carried out by
adding a constant correction to each mean Lagrangian radius such as to
bring the numerical data and the theoretical radius into optimal
agreement at time $t = 0$, or for the average of the values at $t = 0$
and $1$, or for the first 10\% of the collapse phase. (In the latter
two cases allowance was made for the evolution of the radii during
this relatively short time interval.)  It may be expected, from our
discussion of the sources of these effects, that they are
$N$-dependent, and indeed we find that the correction to the innermost
Lagrangian radius generally decreases with increasing $N$, and is
smaller (in absolute terms) at larger radii. In what follows we shall
use these adjusted Lagrangian radii for a comparison of the various
models, though we continue to use Fig.4 to illustrate the discussion.

Now we return to the question of how the time scale for evolution
varies with $N$. Fig.5 illustrates, for two particular values of $N$,
one way in which this can be done.  It shows one curve
for each Lagrangian radius, and was constructed as follows. At each
value of $t$ in the $500$-body models (which is the abscissa in Fig.5)
the mean Lagrangian radius was computed, and then by interpolation we
determined the corresponding time at which this mean value of the
radius was reached in the $2000$-body models. (In fact there may be
more than one such time, because statistical fluctuations result in
evolution of the radius which is not quite monotonic.  We chose the
first such time, and checked that our results are insensitive to this
choice.) According to the theory of relaxation the ratio of these
times, which is the scale factor plotted in Fig.5, should be
$S_f=N_2\ln(\gamma
N_1)/ N_1\ln(\gamma N_2)$, where in this case $N_1 = 500$ and $N_2 =
2000$.

The results for all combinations of $N_1$ and $N_2$ show remarkable
consistency. The values of $S_f$ for the innermost Lagrangian radii
($1$ - $10\%$) and outermost Lagrangian radii ($75$ and $90\%$) are
very close together.  On the other hand the $20$ - $40\%$ Lagrangian
radii lie below and the $50\%$ radius lies above; this means that
these intermediate radii evolve too fast in models with smaller $N$.
The explanation for this may be connected with the fact that binaries
start to influence the evolution relatively earlier in core collapse
in low-$N$ systems than for larger $N$ (cf. Giersz \& Spurzem 1993 and
also Paper 2 in this series).  The effect of binaries would not be
confined to intermediate radii, but that is where their influence
might be most noticeable in this kind of analysis, because these radii
evolve less quickly than those further in and further out.

Fig.5 shows clear evidence of the Coulomb logarithm in the relaxation
time: without it we would predict $S_f = 4$ for the relevant values of
$N_1$ and $N_2$.  One would like to be able to use this data for an
empirical determination of $\gamma$, but this is very difficult
because of the weak predicted dependence of $S_f$ on $\gamma$:
fluctuations in $S_f$ give unacceptably large fluctuations in
$\gamma$.  To determine $\gamma$, therefore, we adopted a somewhat
more indirect method, which we now describe.

Two-body relaxation theory can be used to study the evolution of stellar
systems in two ways.  One is the gas model (Hachisu \& Sugimoto 1978,
Lynden-Bell \& Eggleton 1980) and the other, which is usually thought to
be more faithful to the picture underlying relaxation theory, is the
Fokker-Planck model (cf. Spitzer 1987). Both models, however, share some
significant simplifying assumptions, including spherical spatial
symmetry, isotropy in the velocity distribution, and no loss of mass by
escape, at least in the simplest formulations of these two models.

In order to use results from these models to evaluate the evolution of
an $N$-body model, it is necessary to relate the time variables in the
different models. Conversion between Fokker-Planck and $N$-body models
depends only on the value of the
coefficient $\gamma$  in the expression
for the relaxation time, i.e.
$$
t_r = {0.065(\sqrt{3}\sigma)^3\over \rho m G^2\ln(\gamma N)} \eqno(1)
$$
(cf. H\'enon 1975, Spitzer 1987), where $\sigma$ is the rms value of
each component of velocity, $\rho$ is the stellar mass density, and
$m$ is the individual stellar mass.
Conversion between gas and N-body models requires in
addition the specification of a dimensionless conductivity coefficient,
denoted by $C$ (Lynden-Bell \& Eggleton 1980).  A number of previous
theoretical estimates are listed in Table 2.

If the time variable of a set of $N$-body models is denoted by $t_N$,
the scaling between this variable and the time variable of a continuum
model was carried out in a manner analogous to the determination of
the scale factor $S_f$ in Fig.5.  For each value of $t_N$ this gives a
single estimate for the scalings which bring the models into
agreement.  A typical result is shown in Fig.6, though since this
illustrates scaling with the Fokker-Planck model the result has been
expressed in terms of the equivalent value of $\gamma$. This case illustrates
a fairly general finding, that the inner radii agree well within
themselves, but that the half-mass and $75\%$ radii are relatively
discrepant.  Also noticeable in the inner radii is a tendency for the
value of
$\gamma$ to decrease as  $t_N$ increases; this is most noticeable in runs with
the smallest $N$, and will be discussed further in \S3.1.3..

We consider first the results for the radii well inside the half-mass
radius.  As can be inferred from Fig.6, individual estimates of
$\gamma$ range widely, especially near the beginning of the evolution.
Elsewhere, however, the range of values is smaller, and roughly
independent of time and radius.  Examination of results for other sets
of calculations also show that it is roughly independent of $N$.  The
great bulk of the values lie in the range $0.07 \leq \gamma \leq
0.14$, and show greater consistency for larger values of $N$ than for
$N = 250$.  For the gas model the scale factor shows similar
fluctuations, but their interpretation is more complicated, as the
scaling depends on two parameters, viz.  $\gamma$ and $C$.

These data can also be used to estimate global values (one for each
value of $N$) for the scalings between the $N$-body models and the two
continuum models. The results were then used to estimate the
$N$-independent parameters $\gamma$ and $C$, with results shown in
Table 2. For the gas models, the estimate of both parameters can be
made from each pair of values of $N$, which was done in order to
investigate the variation of these parameters.  The values presented
in Table 2 are, however, obtained from a least squares fit to the
results for all $N$.  The Fokker-Planck data yield one estimate of
$\gamma$ for each $N$, but again the value in Table 2 is a mean.
(Note, however, that the process is iterative, as the determination of
the adjustment to the Lagrangian radii already requires a knowledge of
the scaling between the $N$-body and Fokker-Planck models, i.e. a
preliminary value of $\gamma$.)

The variations about these means are quite modest.  For the
Fokker-Planck determination of $\gamma$, results for the four values of
$N$ range from about $0.108$ to about $0.120$.  For the gas model the
range of values of $C$ is also quite small, from about $0.099$ to about
$0.106$, but the range for $\gamma$ is greater: $0.104$ to $0.138$.
Note that these variations are correlated, larger values of $\gamma$
corresponding to (and partially compensating) smaller values of $C$.
Despite these variations, the consistency is unexpected, especially when
it is considered that $\gamma$ is a coefficient in the argument of a
logarithm which is usually regarded as ``slowly varying''.  An
indication of the overall success of the resulting parameters, at
least for the inner radii, is
provided by Fig.7, which for one value of $N$ shows the comparison
between the gas, Fokker-Planck and $N$-body models.

It should be stressed again that our empirically determined values of
$\gamma$ and $C$ are based mainly on data for the innermost Lagrangian
radii.

Also shown in Table 2 are some earlier theoretical estimates of the
parameters $\gamma$ and $C$.  It is interesting to point out that one of
H\'enon's values of $\gamma$ stemmed from a close reexamination of
Chandrasekhar's theory of relaxation, which he undertook in order to
improve a previous comparison between the results of $N$-body and
Fokker-Planck models (Aarseth, H\'enon \& Wielen 1974). It is only
slightly greater than our value.

\medskip
{\bf 3.1.3 \quad Differences between $N$-body and continuum results}
\medskip

We now discuss the discrepant results in Fig.6 (and related data),
beginning with the half-mass radius.  It is well known that the
half-mass radius is nearly constant in the collapse of a Plummer
model, which can be understood in terms of energy conservation and the
empirical fact that the half-mass radius, expressed as a fraction of
the virial radius, is not sensitive to the model under consideration
(Spitzer 1987).  The implication of this near constancy, however, is
that the scale factor from which $\gamma$ is computed is very
sensitive to fluctuations in the $N$-body data.  Nevertheless the
difference between the behaviour of the half-mass radius and that of
the inner radii is systematic, in the sense that the $r_h$ grows too
quickly in the continuum models (Fig.7), a fact which was already noted in a
previous small-scale statistical study of $100$-body systems (Heggie
1991). As was mentioned in \S3.1.2 similar
behaviour is exhibited by the $20$ and $75\%$ Lagrangian radii, and
this already becomes  visible soon after the start of the evolution.

A possible reason for this is the development of anisotropy in the
$N$-body models, which is discussed further in \S3.2 below and in much
detail by Giersz \& Spurzem (1993).  Their finding is that the evolution
of $r_h$ is not even very well described by any reasonable anisotropic
gaseous model. For the half mass radius the best agreement with the N-body
data is provided by the anisotropic Fokker-Planck models of Stod\'o\l kiewicz
(1982) who uses a Monte-Carlo technique. Since the only relevant published
data from these models is diagrammatic, a quantitative measure of the
agreement is not yet possible.

Another factor which is neglected
in the continuum models is the change in mass ($M$) and energy ($E$)
of the system caused by escape (cf. \S3.3 below).  We have corrected
the values of $r_h$ in the gas models by using the actual values of
mass and energy in the $N$-body models, assuming that $r_h\propto
M^2/E$.  The correction has a noticeable effect, but does not
substantially improve the agreement between the two models.

It is easily seen in Fig.7 that the discrepancy is worse in the gas
model than in the Fokker-Planck model, the former evolving more slowly
beyond the 10\% radius. This has nothing to do with the $N$-body
models, of course, but is a deficiency in the gas model. In effect it
means that the value of $C$ must increase slightly with radius,
reinforcing the fact that the values in Table 2 refer to the innermost
radii.

The other noticeable discrepant feature in Fig.6 is the slight
decrease in the scaling factors with time. It is related to a
systematic discrepancy between the evolution of the continuum and
$N$-body radii which is visible particularly clearly in Fig.5.  This
trend is qualitatively consistent with two possible explanations, both
of which are of dynamical interest.  One is the assumption, discussed
by Spitzer (1987), that, for the core, the argument of the Coulomb
logarithm in eq.(1) should be proportional to $N_c$, the number of
stars in the core, rather than $N$ (cf.  Table 2).  We discuss this
possibility further below.  The other possible explanation is the
growing activity of binary stars, which, in such relatively small
models, begins to influence the collapse of the core well before it
brings the collapse to a close.  (This process is not modeled at all
in the Fokker-Planck and gas models used for this comparison.) We
shall discuss this in the next paper in this series, which analyses
the behaviour of binaries throughout the evolution, including the
phase beyond core collapse. As a precaution, however, in the
determination of $C$ and $\gamma$ we have discarded data at times when
binaries have absorbed more than $1\%$ of the total energy of the
system.

In order to investigate the effect of a variable Coulomb logarithm we
have repeated the computation of gas models but have chosen for the
Coulomb logarithm the expression $\ln(\gamma N_c)$ if $N(r)<N_c$ and $\ln
(\gamma N(r))$ otherwise, where $N_c$ is the number of stars in the core
(Spitzer 1987, p.149) and $N(r)$ is the number of stars within radius
$r$. The value of $\gamma$ chosen was $0.1$ (cf. Table 2), and the
computation carried out for $N=250$, for which the problems with fitting
the Lagrangian radii by results of gas computations seemed most
difficult. The rationale for the choice of Coulomb logarithm starts from
the fact that the argument of the logarithm should be the ratio of the
largest effective impact parameter to the $90^\circ$ deflection
distance.  The latter, which depends on the velocity dispersion, varies
rather slowly throughout the system.  The former may be taken as being
comparable with the radius of that part of the system where the density
is not much smaller than the local value. Within the core this distance
will be of order $r_c$, the core radius, and outside the core it is of
order the radius $r$ itself.  In a nearly isothermal halo (outside the
core) $r\propto N(r)$ approximately.

The result of this calculation showed an overall slowing of the
evolution, since the modification of the Coulomb logarithm from
$\ln(\gamma N)$ lengthens the relaxation time.  But the effect decreases
as one moves to larger radii.  Compared with the $N$-body models with
$N=250$, it was found that the gas model with variable Coulomb logarithm
gave results for the scale factor (between time variables for the gas
and $N$-body models) which were a little more consistent
for different radii (but still excluding the half-mass radius).
On the other hand the
time-dependence of the scale factor was virtually unimproved, which
suggests that this has some other explanation, such as binary
evolution.

In conclusion, there is some evidence favoring use of a variable
Coulomb logarithm.  In addition, the comparison between $N$-body and
Fokker-Planck data gives some indication that $\gamma$ may increase
slightly with increasing radius, which is qualitatively consistent
with our prescription for a variable Coulomb logarithm . Unfortunately
this result cannot be conclusive because of the large fluctuations in
the values of $\gamma$ (cf. Fig.6).

\medskip
{\bf 3.2  \quad The velocity distribution}
\medskip

Fig.8 shows some typical results for the velocity dispersions (radial
and tangential, or the sum of these).  The general increase with
time in the inner parts of the cluster (Fig.8a) is to be expected from
core collapse, and the results from smaller $N$ models show a very
similar trend with greater noise.  The illustrated comparison with the
results of gas calculations is satisfactory until late in the core
collapse phase, and there is even a suggestion that the $N$-body
models exhibit the initial cooling, which is also found in
Fokker-Planck models (cf. Cohn 1980, Fig.7).  There is no discernible
anisotropy in the innermost shell, and slight evidence (seen in a
graph of the quantity $\langle v_t^2\rangle/\langle v_r^2\rangle$,
which is not, however, reproduced here) between the Lagrangian radii
for $20$ and $50\%$ of the mass.  Anisotropy clearly grows, however,
in the outer shells (Fig.8b).  There is a slight decline in the
radial velocity dispersion, but the tangential dispersion decreases
much more rapidly.  The overall downward trend of the mean velocity
dispersion agrees well with the predictions of an isotropic gas model
and is associated with the general expansion of this zone, but the
radial dispersion is maintained by the ejection of stars from the
inner parts of each system.  The growth of anisotropy (measured by
$A=2 - \langle v_t^2\rangle/\langle v_r^2\rangle$) with time is quite
linear, and it reaches values between $0.9$ and $1.1$ (the larger
value corresponding to the smallest value of $N$) at the end of core
collapse. Detailed comparisons with {\sl anisotropic} gaseous models
are presented in Giersz \& Spurzem (1993).

\medskip
{\bf 3.3 \quad Escape}
\medskip

{\bf 3.3.1 \quad Review of previous work}
\medskip
The rate at which stars escape from an isolated clusters of equal stars
is a problem with a long history (Table 3). Early theoretical estimates
(Ambartsumian 1938, Chandrasekhar 1943) were based on the theory of
relaxation, and led to predictions of a fractional escape rate per
relaxation time which is independent of $N$. This translates to an
escape rate per crossing time which is proportional to $\log N$.
Relaxation being understood as a diffusive process, H\'enon (1961) (who
also lists some other theoretical estimates of the escape rate) then pointed
out the difficulty of diffusing stars across the energy of escape, when
their periods should become extremely long. He therefore gave an
estimate of escape from two-body interactions, though the result depends
on the assumed distribution of velocities; it is
$$
{dN\over dt} = - {256\sqrt{2}\pi^4G^2m^2\over3}\int_0^\infty r^2 dr
\int\int {(\varepsilon + \varepsilon^\prime -
\phi)^{3/2}\over\varepsilon^2}f(\varepsilon)f(\varepsilon^\prime)d\varepsilon
d\varepsilon^\prime, \eqno(2)
$$
where $m$ is the individual stellar mass, $r$ is
the distance from the centre of the cluster, $f(\varepsilon)$ is the
phase-space density expressed as a function of the specific energy
$\varepsilon$, $\phi(r)$ is the potential, and the integration is taken
over the range $\varepsilon < 0, ~\varepsilon^\prime < 0, ~\varepsilon +
\varepsilon^\prime > \phi$.
H\'enon's point of view was further modified by Spitzer \& Shapiro (1972), who
pointed out that the distribution itself evolves on a relaxation time
scale, and then a star which has diffused to energies a little below the
escape limit can escape in a single two-body encounter in the core.
Some previous numerical estimates for the escape rate are also summarised in
Table 3.
\medskip
{\bf 3.3.2 \quad Results from the present $N$-body simulations}
\medskip

Escaping stars in our $N$-body models are identified according to
the following definition.  Their energy (computed in the rest frame of
the centre of mass of the entire system) must be positive, and their
distance from the density centre must exceed a boundary radius which is
chosen as $10$ or $20$ times the half-mass radius. The results presented
in this paper are for boundary radius equal to $20$ times $r_h$,
except where  stated in \S3.3.3.

Two typical sets of results are shown (along with a theoretical comparison,
which we discuss below) in Fig.9.  Note that these are means over many
runs. We have checked that the mean agrees quite well with the median
number of escapers, but there is a wide range about the mean.  For $N =
250$, for example (not shown), the mean number of escapers at the end of core
collapse is approximately 4.0, but in individual models may range from $0$
to $9$.  Note the clear increase in the rate of escape with time; such an
increase was already noted in the Monte Carlo models of Spitzer \& Shull
(1972), who found that it increased by a factor of about four during the
course of core collapse. We found that it increased by a factor of about 6 for
$N=250$. For higher $N$ the increase is even larger (Fig.10).

For a first comparison between our own results and those of theory we
have adopted a synthesis of the general theoretical result of H\'enon
with the ideas of Spitzer \& Shapiro.  What we have done is to apply
H\'enon's result to the time-dependent distribution function given by
the (isotropic) Fokker-Planck model.   Results of such a comparison
are shown in Fig.9, which shows that the agreement is rather poor,
except initially.  Up to the end of core collapse the mean number of
escapers in the $N$-body models exceeds the theoretical prediction by
a factor of about $1.8$ and $2.8$ for $N=500$ and $1000$,
respectively.  The results for other values of $N$ are qualitatively
similar, and the agreement at early times is clearer in the larger
models than smaller one (as we can see in Fig.9b).  For all models
(but particularly for this with low $N$) the comparison is complicated by
the fact that stars are only deemed to escape after they have reached
a sufficiently large radius. This depresses the escape rate at very
early times (cf. also Fig.10), and makes the number of escapers
depend on the definition of the boundary radius.

The disagreement between $N$-body and Fokker-Planck results is not surprising,
because eq.(2) is based on the assumption that the distribution of
velocities is isotropic.  This is true of all our theoretical
comparisons, but it is especially critical in this case, because escape
tends to take place for stars which are already loosely bound and yet
have sufficiently small angular momenta to pass through the core.
In addition to anisotropy, which we discuss further below,
further possible explanations of this
discrepancy are explored
in Paper II.

Fig.11 shows the situation for the energy carried off by escapers, i.e.
their asymptotic kinetic energy at infinity.  What is plotted here is,
as usual, the mean value over all the simulations at this value of $N$.
Use of the median would be particularly unsuitable for escaper
statistics, as a substantial fraction of cases may have no escapers for
a significant part of the core collapse phase.

In Fig.11 the theoretical result is based on a general expression given
by H\'enon (1969) for isotropic distribution functions, and evaluated by
him for a Plummer model with unequal masses.  As before we applied it to
the evolving distribution function produced by an isotropic
Fokker-Planck code.  Again there is agreement at early times, but a
faster rise throughout much of core collapse.  Even more dramatic is the
abrupt rise towards the close of core collapse, and it is natural to
interpret this in terms of escapers produced in three-body encounters.
Again discussion of this process is relegated to Paper II in this
series.

\medskip
{\bf 3.3.3 \quad The effect of anisotropy}
\medskip

In order to understand the influence of anisotropy on the rate of escape
it would be desirable to calculate the evolution of the distribution
function using an anisotropic model, and then repeat our computation of
the escape rate using a suitable anisotropic generalisation of H\'enon's
formula, eq.  (2).  In fact anisotropic gas models have been developed
by several authors (Larson 1970, Bettwieser 1983, Louis and Spurzem 1991),
and a recent version of such a code is compared with the results of our
and other $N$-body data in Giersz \& Spurzem (1993).  Unfortunately,
however, the gas model does not directly yield the distribution function
in phase space, and so these results cannot be applied to the discussion
of the escape rate.

Since the effect of anisotropy on escape appears to be quite unknown, we have
formulated a model problem which should provide a semi-quantitative
guide.  What we have done is
to use a Monte Carlo method to compute the escape rate in a sequence of
anisotropic models.  The sequence we chose was devised by Dejonghe
(1987), and has the benefit that all the models in the sequence have the
same spatial density distribution as in Plummer's model.  This allows us
to compare our results with H\'enon's well known results for this case
(H\'enon 1961, 1969).  More importantly, the models have the same density
distribution as our initial conditions; thus the study of this sequence
of models allows us to determine the effects of anisotropy, while our
study of the Fokker-Planck model allows us to study the effect of
dynamical evolution.  Dejonghe's models have one scale-free  free
parameter, $q$, which we
took in the range from $0$ (Plummer's model) to $1.5$.  It is related
to the anisotropy $A$ defined in \S3.2 by
$$
q  = A(1+1/r^2).\eqno(3)
$$

In order to compute the escape rate the following procedure was used.
First, the radius $r$ is selected from a probability density function
proportional to $4\pi r^2 n(r)^2$, where $n$ is the number-density.
Next, the velocities of two stars are selected from the local
distribution of velocities.  Then the impact parameter $p$ is chosen
with probability density proportional to $p$ up to a maximum value
$p_{max}$, and an angle which
determines the relative orientation of the two stars at the moment when
they would be at their minimum separation during an encounter, if their
paths were undeflected by it.  Then it is possible to
determine whether either star has a speed, after the encounter, above
the local escape speed.  Then the escape rate is the average value of
$v$, the relative speed of the two stars before the encounter, suitably
normalised. The resulting value is subject to statistical uncertainty,
and depends on the choice of $p_{max}$.  Experimentation with different
values, and different numbers of trials, allowed us to determine
satisfactory choices.

The results of our calculations are shown in Figs.12 and 13.  Units are
standard, i.e. those in which $M = 1,~G = 1$ and $E=-1/4$. For $q = 0$
the results may be compared with those given by H\'enon, i.e. $\dot N =
0.00942$ and $\dot E = 0.000741$. Evidently the presence of anisotropy can lead
to a large increase in the rate of escape and the flux of energy carried
off by escaping stars.

In applying these results to a comparison with our $N$-body data we have
made two rather arbitrary assumptions.  One is that the effects of
evolution (discussed in the previous section) and anisotropy can be
combined multiplicatively.  The second is that we can estimate the
effect of anisotropy by  using
the analytical (Plummer-like)
model which has a value of $q$ given by eq.(3), in which $A$ is taken
to be the
average anisotropy in the outer half of the cluster and $r$ is the 75\%
Lagrangian radius.  When this is done it is
found that the predicted number and energy of escapers agree with
those from the $N$-body simulations quite well as can be seen in
Figs.9 and 11.
The agreement is particularly good for $N \ge 1000$. For smaller $N$ the
predicted number and energy of escapers are larger then those obtained from
the $N$-body data, and for $N=250$ the discrepancy is about $40\%$ of
the $N$-body
value. The largest discrepancy that  we have noticed is the
energy flux at late times in core collapse; as already mentioned it rises
rather suddenly as
the end of core collapse is approached.
It is probable that the increased
contribution from three-body escape events (i.e. those associated with
binaries; cf. Paper II) is beginning to make a significant contribution.
Another explanation which we cannot rule out is the inappropriateness of
one or other of the assumptions mentioned earlier in this paragraph,
though the abruptness of the rise suggests otherwise.

As we mentioned before the $N$-body data depend on the
definition of the boundary radius. If the boundary radius is reduced
to $10r_h$ the predicted values agree with those from $N$-body models
to better than $25\%$ throughout most of core collapse. But now, for
all values of $N$ that we have studied, they are smaller than the
$N$-body values.  Therefore we can expect that in the case of
instantaneous removal of escapers from $N$-body models (which is in
effect what happens in the theoretical model) the discrepancies
between computed and predicted values will be larger than for the
model with boundary radius $10r_h$.  But at any rate, our conclusion
is that anisotropy has a large enough effect on the escape rate that
it may be quite enough to reconcile theoretical predictions with our
numerical results.

\bigskip

{\bf 4 \quad DISCUSSION AND CONCLUSIONS}
\medskip

We have computed large numbers ($\sim 200$) of isolated $N$-body
simulations with $N$ in the range from $250$ to $2000$.  All stars have
equal masses, and initially the systems are in dynamic equilibrium. The
purpose of these studies is to improve the quantitative results of such
computations by statistically combining data from many simulations.

We find that the evolution of the spatial distribution of stars, up to
the end of core collapse, is quantitatively quite consistent with the
theory of relaxation.  Indeed we have been able to estimate a rather
reliable value for the Coulomb logarithm.  Small deviations from the
predictions of isotropic models based on the theory of relaxation are
present at small radii (where they are probably associated with
incipient binary activity) and at the $50$ and $75\%$ Lagrangian radii
(where anisotropy of the velocity dispersion may be responsible).

Inside the radius containing half the mass of the cluster, the
evolution of the velocity dispersion is also consistent with
simplified models in which the distribution of velocities is assumed
to be isotropic, but there is strong growth of anisotropy in the outer
half of the mass.  A quantitative comparison with the anisotropy
predicted by a simplified theory of relaxation is presented elsewhere
(Giersz \& Spurzem 1993).

The rate of escape of stars is roughly consistent with a hybrid model
based on existing theory for escape in evolving, isotropic systems, with
a somewhat {\sl ad hoc} but important modification due to anisotropy.
The escape is assumed to take place by two-body interactions from
stars whose distribution function slowly changes in accordance with
the usual theory of relaxation.

\bigskip
\vfill\eject

{\bf ACKNOWLEDGEMENT}
\smallskip

{\parindent=0pt
We are indebted to J.  Blair-Fish of Edinburgh Parallel Computer Centre
for much help in launching this program on maxwell, the Edinburgh
Concurrent Vector Supercomputer.  We are grateful also to the Centre for
substantial allocation of time on this machine in its early days, and to
the Faculty of Science of the University of Edinburgh for subsequent
allocations of time.  Other computations have been carried out with the
Edinburgh Computing Surface, a transputer array operated by Edinburgh
University Computing Service on behalf of the Edinburgh Parallel
Computing Centre, to whom we offer further thanks.  We are grateful to
B.R.P. Murdoch of EUCS for allowing one of us (DCH) to conduct one
series of simulations on a Dec Alpha machine on field test.
We would like
to thank H.-M.  Lee and H.N.  Cohn for kindly making available to us a
version of their isotropic Fokker-Planck code for the purpose of
comparison with the $N$-body results.  Conversations with R.  Spurzem
have been very fruitful in the comparison of the $N$-body and continuum
models. We thank P. Hut for his comments on a previous draft of the paper.
This research has been supported by a grant (number GR/G04820) of the
U.K.  Science and Engineering Research Council. }
\vfill\eject
\bigskip
{\bf REFERENCES}
\medskip
\leftskip=0.0 truein
\parindent=0.0 truein
Aarseth S.J., 1985, in Brackbill J.U. \& Cohen B.I., eds,
Multiple Time Scales, Academic Press, p.377

Aarseth S.J. \& Heggie D.C., 1992, MNRAS, 257, 513

Aarseth S.J., H\'enon M. \& Wielen R., 1974, A\&A, 37, 183

Aarseth S.J. \& Lecar M., 1975, Ann. Rev. Astron. Ap, 13, p.1

Ambartsumian V.A., 1938, Uch. Zap. L.G.U, 22, 19; transl. in Goodman J.
\& Hut P., eds, Dynamics of Star Clusters, Reidel:Dordrecht, p.521

Barnes J. \& Hut P., 1986, Nature, 324, 446

Bettwieser E., 1983, MNRAS, 203, 811

Casertano S., 1985. in Goodman J. \& Hut P., eds, Dynamics of Star Clusters,
Reidel: Dordrecht, p.305

Casertano S. \& Hut, P., 1985, ApJ, 298, 80

Chandrasekhar S., 1942, Principles of Stellar Dynamics, University of Chicago
Press; reprinted Dover, New York, 1960

Chandrasekhar S., 1943, ApJ, 98, 54

Chernoff D.F. \& Weinberg M.D., 1990, ApJ, 351, 121

Cohn H., 1980, ApJ, 242, 765

Dejonghe H., 1987, MNRAS, 224, 13

Giersz M. \& Spurzem R., 1993, MNRAS, to be submitted

Goodman J., 1987, ApJ, 313, 576

Goodman J., Heggie D.C. \& Hut P., 1993, ApJ, in press

Hachisu I. \& Sugimoto D., 1978, Prog. Theo. Physics, 60, 123

Heggie D.C., 1973, in Tapley B.D. \& Szebehely V., eds, Recent Advances
in Dynamical Astronomy, Reidel:Dordrecht, p.34

Heggie D.C., 1985, in Goodman J. \& Hut P., eds, Dynamics of Star
Clusters, Reidel:Dordrecht, p.139

Heggie D.C., 1989, in Merritt D., ed, Dynamics of Dense Stellar Systems,
Cambridge University Press, p.195

Heggie D.C., 1991, in  Roy A.E., ed, Predictability, Stability and Chaos
in $N$-Body Dynamical Systems, Plenum, New York, p.47

Heggie D.C.  \& Mathieu R.D., 1986, in Hut P. \& McMillan S.L.W., eds,
The Use of Supercomputers in Stellar Dynamics, Springer-Verlag, Berlin, p.233

Heggie D.C. \& Stevenson D., 1988, MNRAS, 230, 223

H\'enon M., 1961, Annales d'Astro\-phys\-ique, 24, 369

H\'enon M., 1969, A\&A 2, 151

H\'enon M., 1975, in Hayli A., ed, Dynamics of Stellar Systems,
Reidel:Dordrecht, p.133

Hut P., Makino J., 1989, Comp. Phys. Rep., 9, 201

Ito T., Ebisuzaki T. Makino J. \& Sugimoto D., 1991, PASJ, 43, 547

Larson R.B., 1970, MNRAS, 147, 323

Louis P.D. \& Spurzem R., 1991, MNRAS, 251, 408

Lynden-Bell D. \& Eggleton P.P., 1980, MNRAS, 191, 483

Makino J. \& Sugimoto D., 1987, PASJ, 39, 589

McMillan S.L.W. \& Aarseth S.J., 1993, ApJ, in press

McMillan S., Casertano S. \& Hut P., 1988,  in Valtonen M., ed,
The Few Body Problem, Kluwer:Dordrecht, p.313

McMillan S., Hut P. \& Makino J., 1990, ApJ, 362, 522

McMillan S.L.W., Hut P. \& Makino J., 1991, ApJ, 372, 111

Spitzer L., Jr., 1987, Dynamical Evolution of Globular Clusters,
Princeton University Press, Princeton

Spitzer L., Jr. \& Shapiro S.L., 1972, ApJ, 173, 529

Spitzer L., Jr. \& Shull J.M., 1972, ApJ, 200, 339

Stod\'o\l kiewicz J.S., 1982, Acta Astron, 32, 63

Sweatman W.L., 1993a, J. Comp. Phys., submitted

Sweatman W.L., 1993b,  MNRAS, 261, 497

von Hoerner, S., 1960, Z. Ap, 50, 184

Wielen R., 1974, in Proc.  1st Europ. Astron. Meeting, Athens, Sep 4-9, 1972,
2, p.326

\leftskip=0pt
\vfill\eject

\centerline{{\bf Table 1}}
\medskip

\centerline{OUTLINE OF THE COMPUTATIONS}
\medskip

$$\vbox{ \settabs \+\quad 2500 \quad & \quad number of cases \quad &
\quad cpu (hrs) \quad & \quad copernicus (Sun ELC) \quad & \cr
\+ \hfill $N$ \hfill & \hfill number of cases \hfill & \hfill cpu (hrs)
\hfill	& \hfill machine \hfill &\cr
\smallskip
\+ \hfill 250 \hfill & \hfill  56 \hfill & \hfill 50 \hfill & \hfill
super \hfill &\cr
\+ \hfill 500 \hfill &  \hfill 56 \hfill & \hfill 200 \hfill & \hfill
super \hfill &\cr
\+ \hfill 1000 \hfill & \hfill 40 \hfill & \hfill 190 \hfill & \hfill
maxwell \hfill& \cr
\+ \hfill 1000 \hfill & \hfill 50 \hfill & \hfill 1200 \hfill & \hfill
fringe \hfill &\cr
\+ \hfill 2000 \hfill & \hfill 16 \hfill & \hfill 1350 \hfill & \hfill
super \hfill &\cr
\+ \hfill 2500 \hfill & \hfill 2 \hfill  & \hfill  13000  \hfill  &\hfill
copernicus (Sun ELC) \hfill& \cr}$$
\bigskip
Note: For parallel machines (super and maxwell) the total number of
processor-hours is the product of columns 2 and 3.
\bigskip
\bigskip
\vfill\eject

\centerline{{\bf Table 2}}
\medskip

\centerline{ESTIMATES OF RELAXATION PARAMETERS}
\medskip
$$\vbox{ \settabs \+ Parameter & $2N_c/N$& This project
(Fokker-Planck model)& number of stars in core &\cr
\+ \qquad Parameter  &		& \hfill Source \hfill & \hfill
Notes \hfill& \cr
\+ \hfill C \hfill & \hfill $\gamma$ \hfill & &	&\cr
\smallskip
\+ \hfill --- \hfill & \hfill 0.25 \hfill & \hfill Ambartsumian (1938)
\hfill & &\cr
\+ \hfill --- \hfill & \hfill 0.35 \hfill & \hfill Chandrasekhar (1942)
\hfill & &\cr
\+ \hfill --- \hfill & \hfill 0.15 \hfill & \hfill H\'enon (1975) \hfill &
\hfill Depends slightly on \hfill&  \cr
\+		&		&		& \hfill velocity
distribution \hfill& \cr
\+ \hfill --- \hfill & \hfill 0.2 \hfill & \hfill McMillan
({\sl pers. comm.}) \hfill &\cr
\+ \hfill --- \hfill & \hfill 0.4 \hfill & \hfill Spitzer (1987) \hfill & &\cr
\+ \hfill --- \hfill & \hfill $2N_c/N$ \hfill & \hfill Spitzer (1987)
\hfill & \hfill For the core; $N_c=$ \hfill &\cr
\+		&		&		& \hfill number of
stars in core \hfill& \cr
\+ \hfill 0.104 \hfill & \hfill --- \hfill & \hfill Heggie \&
Stevenson (1988), \hfill &\cr
\+		&		& \hfill Heggie (1985), Goodman (1987)
\hfill &\cr
\+ \hfill 0.104 \hfill 	& \hfill 0.11 \hfill & \hfill This project
(gas model) \hfill &	\cr
\+ \hfill --- \hfill & \hfill 0.115 \hfill & \hfill This project
(F-P model) \hfill & \hfill \quad \hfill &\cr}$$
\bigskip
\bigskip
\vfill\eject
\centerline{{\bf Table 3}}
\medskip

\centerline{THE ESCAPE RATE}
\medskip

$$\vbox{\settabs \+ Spitzer \& Shull (1972) &\quad $-t_{cr}dN/dt$\quad &
Relaxation; $ln\Lambda=$ Coulomb logarithm &\cr
\+ \hfill Source \hfill & \hfill $-t_{cr}dN/dt$ \hfill & \hfill Notes
\hfill &	\cr
\smallskip
\+ \hfill Ambartsumian(1938) \hfill & \hfill $0.16\ln\Lambda$ \hfill 	&
\hfill Relaxation; $\ln\Lambda =$ Coulomb logarithm \hfill &	\cr
\+ \hfill Chandrasekhar(1943) \hfill & \hfill $0.33\ln\Lambda$ \hfill 	&
\hfill Relaxation and dynamical friction \hfill &\cr
\+ \hfill H\'enon(1961) \hfill 	& \hfill  $0.027$ \hfill & \hfill Plummer
model; 2-body encounters \hfill &\cr
\+ \hfill Spitzer \& Shull (1972) \hfill & \hfill  $0.06\ln\Lambda$
\hfill & \hfill Mean value in core collapse \hfill &\cr
\+ \hfill Wielen (1974) \hfill 	& \hfill $\simeq0.1$ \hfill & \hfill $50
\le N\le 250$ \hfill &	\cr
\+ \hfill Aarseth \& Lecar (1975) \hfill & \hfill  $0.2$-$0.3$ \hfill &
\hfill Plummer model ($N = 250$) \hfill &\cr
\+ \hfill Stod\'o\l kiewicz (1982) \hfill & \hfill $0.3$ \hfill &
\hfill Maen value in core collapse \hfill &\cr
\+ \hfill Heggie (1991)	 \hfill & \hfill $0.09$	 \hfill & \hfill $N = 100$
\hfill &\cr}$$
\bigskip

\vfill\eject

{\bf FIGURE CAPTIONS}
\medskip

{\bf Fig.1} Statistics of the $5\%$ Lagrangian radius for $56$ cases
with $N=250$.  The curves correspond to the minimum, median, mean and
maximum at each time.

{\bf Fig.2} Mean Lagrangian radii for (a) $56$ cases with $N = 250$,
compared with (b) a single typical case.

{\bf Fig.3} Mean Lagrangian radii for all series, for a mass fraction
of $2\%$. For a Plummer model the value is $0.166$.

{\bf Fig.4} Comparison between $N$-body, gas (IGM) and Fokker-Planck
(F-P) models for the $1\%$ Lagrangian radius and $N = 250$. The time
scales of the continuum models have been scaled empirically to provide
a reasonable global fit; the values used were $C = 0.104,~\gamma=0.11$
(cf.Table 2).

{\bf Fig.5} The scale factor in time which brings $N=500$ and $N=2000$
models into agreement. All ten Lagrangian radii are shown. The
discrepant ones are $20\% - 40\%$ radii (below the others) and
half-mass radius (above the others).  The thick line shows scale
factor for $\gamma =0.11$.  The abscissa is the time variable in the
500-body calculations.

{\bf Fig.6} The value of $\gamma$ which brings Fokker-Planck and
$N$-body results into agreement at each time, for $N = 500$.  The
discrepant Lagrangian radii (below the others) are the $50\%$ and
$75\%$ Lagrangian radii.

{\bf Fig.7} Comparison between the gas (IGM), Fokker-Planck (F-P) and
$N$-body models with $7$ Lagrangian radii for $N = 1000$, to indicate
the overall success of our fitted values of the relaxation parameters
$\gamma$ and $C$ (cf. Table 2).

{\bf Fig.8} Velocity dispersions for $N$-body models with $N = 1000$.
(a) Total velocity dispersion between Lagrangian radii for $1\%$ and
$2\%$ of the mass, compared with a gas model.  (b) Radial and
transverse dispersions between Lagrangian radii for $75\%$ and $90\%$.

{\bf Fig.9} Number of escapers as a function of time for different sets
of models.  F-P-I - isotropic Fokker-Planck model, P-A-C - anisotropic
Plummer model with evolution (see text for discussion).  (a) for
$N = 500$, (b) for $N = 1000$.

{\bf Fig.10} The escape rate as a function of time for $N = 1000$. The
squares represent data given by Spitzer \& Shull (1972) for a Monte-Carlo
model. Their time unit has been scaled to $N$-body time for $N =1000$.
The triangle represents data obtained by Stod\'o\l kiewicz (1982) for a
Monte-Carlo model. The dot represents the result given by H\'enon (1961)
for a Plummer model. The labels $10 r_h$ and $20 r_h$ correspond to the
boundary radii chosen in different sets of $N$-body simulations (see text for
discussion).

{\bf Fig.11} Energy of escapers as in Fig.9.

{\bf Fig.12} Escape rate for an analytic sequence of anisotropic
models (the aniso\-tropy being determined by $q$).  Error bars are
estimated 1-$\sigma$ confidence limits. The dashed line and the solid
curve are linear and quadratic fits of the form $a+bq$ and $a+bq+cq^2$
respectively.  The linear fit is adequate for all except the lowest
value of $q$.

{\bf Fig.13} Rate at which energy is carried off by escapers in a
sequence of analytical models. For other details see the caption to
Fig.12.
\bye